\documentstyle[tighten,preprint,amsfonts,%
epsf,eqsecnum,aps,prd]{revtex}
\begin{document}
\draft

\preprint{UBC-SW-96-1}
\title{The Generalized Hartle-Hawking Initial State:
Quantum Field Theory on Einstein Conifolds\\}
\author{Kristin Schleich and Donald M. Witt\\}
\address{
Department of Physics and Astronomy\\
University of British Columbia\\
Vancouver, BC V6T 1Z1 Canada\\
}
%%\date{\today}
\maketitle
\begin{abstract}
Recent arguments have indicated that the sum over histories formulation of 
quantum amplitudes for gravity should include sums over conifolds, a  set of 
histories with more general topology than that of manifolds. This paper 
addresses the  consequences of conifold histories in gravitational functional 
integrals that also include scalar fields. This study will be carried out 
explicitly for the generalized Hartle-Hawking initial state, that is the 
Hartle-Hawking initial state generalized to a sum over conifolds. In the 
perturbative limit of the semiclassical approximation to the generalized 
Hartle-Hawking state, one finds that  quantum field theory on Einstein 
conifolds is recovered. In particular, the quantum field theory of a scalar 
field on de Sitter spacetime with $RP^3$ spatial topology is derived from the 
generalized Hartle-Hawking initial state in this approximation. This derivation 
is carried out for a scalar field of arbitrary  mass and scalar curvature 
coupling. Additionally, the generalized Hartle-Hawking boundary condition 
produces a state that is not identical to but corresponds to  the Bunch-Davies 
vacuum on $RP^3$ de Sitter spacetime. This result cannot be obtained from the 
original Hartle-Hawking state formulated as a sum over manifolds as there is no 
Einstein manifold with round $RP^3$ boundary.

\end{abstract}

\narrowtext
\section{Introduction\\}
\label{sec:level1}

The sum over histories formulation of quantum amplitudes has been a useful tool 
in the study of quantum gravity, both in formal expressions and in concrete 
calculations. In particular, the Hartle and Hawking proposal for the initial 
state of the universe \cite{hh,h2}, formulated in terms of such a sum, has 
provided a starting point for the study of how the quantum mechanics of the 
early universe leads to its current observed form. This initial state is 
constructed in terms of a sum over regular geometries and field configurations 
on compact manifolds. However, recent arguments by the authors indicate that 
the sum over manifolds used in Euclidean functional integrals for Einstein 
gravity such as the Hartle-Hawking initial state should be extended to a sum 
over a set of more general topological spaces, called conifolds \cite{I}. A 
brief summary of these arguments is the following: First, from knowledge of the 
properties of path integrals in rigorous field theory,  a rigorous definition
of the space of histories for gravity  is anticipated to include 
nondifferentiable geometries. Given this observation, it is natural to also 
consider whether more general topological spaces than those of the classical 
theory should be included in the space of histories. Motivation for doing so can 
be found in the semiclassical analysis of the Hartle-Hawking initial state; one 
can show that there are nonmanifold stationary points of the Euclidean action
for certain 3-manifold boundaries. Moreover, these nonmanifold stationary points 
actually arise as the limit of a sequence of almost stationary geometries on 
manifolds.  In fact, such nonmanifold stationary points  are boundaries of the 
space of Einstein metrics on manifolds. These stationary points are metrically 
complete and can thus be considered regular geometries albeit on a more general 
set of topological spaces. These properties  provide a compelling argument that
such stationary points should be included in semiclassical evaluations of the 
Hartle-Hawking initial state. It then follows that such nonmanifold histories 
should be included in the space of histories. Further considerations, discussed 
in detail in \cite{I} and\cite{II}, lead to the proposal of a particular set of 
nonmanifold histories, called conifolds. 

Some key consequences of formulating  generalized functional integrals for
gravity in terms of a sum over conifolds such as the generalized Hartle-Hawking 
initial state,  were addressed in \cite{I} and \cite{II}. However, these papers 
did not address the more specific issues in formulating these generalized 
functional integrals for the case where scalar fields are also present. In 
particular, an interesting feature of the Hartle-Hawking initial state when 
formulated as a sum over manifolds is that quantum field theory in curved space 
can be recovered in a special limit of its semiclassical approximation 
\cite{hartle}. Moreover, a virtue of this initial state is that it selects a 
unique vacuum state for the matter fields;  as discussed by Hartle 
\cite{hartle}, the Hartle-Hawking initial state for massless conformally coupled 
field perturbations on a round three sphere boundary yields the Bunch-Davies 
vacuum. Additionally, Laflamme \cite{laflamme} showed that similar results hold 
for the case of a massive, minimally coupled scalar field. This nice choice of 
vacuum arises from the requirement that the field configurations are regular 
everywhere on the extrema of the Euclidean action, Euclidean  de Sitter. Thus 
the questions arise; is there a natural notion of regularity for scalar fields 
on conifolds? If so, does this notion result in a unique choice of vacuum for 
the scalar field?

This paper will analyze these questions and formulate the generalized 
Hartle-Hawking initial state for the case of a scalar field with arbitrary mass 
and scalar curvature coupling. Section \ref{section:prelim} will begin with a 
summary  of the essentials of the topology  and geometry of conifolds. Next the 
generalized Hartle-Hawking initial state will be formulated in terms of a sum 
over conifolds. Then the relationship between quantum field theory in curved 
space and the semiclassical approximation to the generalized Hartle-Hawking 
initial state will be delineated. It will be seen that there is a natural notion 
of regularity for scalar fields on conifolds and that this regularity does imply 
a well defined semiclassical  generalized Hartle-Hawking initial state in the 
Euclidean sector. Section \ref{section:pm} will illustrate these properties for 
a particularly interesting case; the semiclassical evaluation of the generalized 
Hartle-Hawking initial state for boundary data consisting of scalar field 
perturbations on ${RP^3}$ with round metric. This semiclassical evaluation
explicitly results in a stationary point that is a conifold. It will be seen 
that the resulting semiclassical wavefunction is not identical to but has a 
direct correspondence to the Bunch-Davies vacuum for $RP^3$ de Sitter for a 
scalar field. This result cannot be obtained from a semiclassical evaluation of 
the original Hartle-Hawking state formulated as a sum over manifolds as there 
is no Einstein manifold with round $RP^3$ boundary.  Section \ref{section:disc} 
will discuss the interesting features of this result. The computation of the 
Bunch-Davies vacuum for a scalar field on $RP^3$ de Sitter spacetime in both 
Fock space and in field representation is given in Appendix \ref{section:bd}.

\section{Generalized Histories for the Hartle-Hawking state}
\label{section:prelim}

A history in the functional integral approach to quantum gravity is specified
by  its topology, smooth structure, and  geometry.  In the original 
Hartle-Hawking initial state, the histories consist of spaces with the topology 
and smooth structure of a smooth compact manifold and a geometry specified on 
that manifold. Generalized histories differ from those in the original 
Hartle-Hawking proposal by having more general topology; the spaces are 
conifolds. 

\subsection{Conifolds}
\label{section:pm1}

A closed n-manifold is a topological space for which the neighborhood of every 
point is homeomorphic to an n-ball \cite{kn}. A closed n-conifold is a more 
general topological space also characterized by the properties of the 
neighborhoods of its points. To precisely define these neighborhoods, one needs 
to define a cone:
\proclaim Definition 2.1.  The cone $C(V)$ of a topological space $V$ is the 
space formed by the cartesian product of the topological space and the unit 
interval $I$ modulo the equivalence relation, $C(V)=(V\times  I)/\sim$ where 
$(v,t) \sim (  v^\prime, t^\prime) \ \ \  t= t^\prime=1$.\par
\noindent Figures \ref{cone} illustrates the construction  of the cone of a
figure eight. 
Figure \ref{crp2} illustrates the cone of the closed manifold $RP^2$. 
Intuitively, one forms the cone by pinching off the top of the product space 
via the implementation of the equivalence relation. Note that this pinching off 
is topological, not geometrical; for example, the cone of $S^{n-1}$ is 
homeomorphic to an n-ball as the equivalence relation in Def. 2.1 defines the 
usual neighborhoods  at all points including the apex of the cone. Thus an 
equivalent definition of a closed n-manifold is a topological space for which 
the neighborhood of every point is homeomorphic to $C(S^{n-1})$.

Now define an open neighborhood $N_{x_0}$ of a point $x_0$ to be
a {\it conical neighborhood} if it is homeomorphic to the interior of 
$C(\Sigma^{n-1}_{x_0})$ with $x_0$ mapped to the apex of the cone where
$\Sigma^{n-1}_{x_0}$ is some closed connected (n-1)-manifold. Then
\proclaim Definition 2.2. A closed n-dimensional conifold $X^n$, $n\ge 2$ is a 
metrizable space such that every point $x_0 \in X^n$ has  a conical 
neighborhood.\par
\noindent It is clear that the set of closed n-conifolds includes all  closed 
n-manifolds. However this set includes more general topological spaces as well 
for $n>2$.\footnote{All two dimensional closed conifolds are 2-manifolds as the 
only closed connected 1-manifold is a circle.}
One can construct a  simple example of a closed n-conifold by taking two cones 
of any closed connected \hbox{(n-1)}-manifold that is not a \hbox{(n-1)}-sphere 
and identifying them at their boundary. Furthermore, one can take two such 
spaces, remove a n-ball from each and identify them along the resulting
\hbox{(n-1)}-sphere boundaries to construct a more general example of a 
n-conifold. In addition, other examples of conifolds are given in 
section 5 of \cite{I}.

A useful  characterization of how  n-conifolds differ from n-manifolds is given 
by the  singular set. The {\it singular set} $S$ is the set of points in the 
conifold $X^n$ whose neighborhoods are not homeomorphic to n-balls. One can 
prove from Def. 2.2 that $S$ consists of discrete points and is countable 
\cite{I}. From this, it can be shown that the space $X^n - S$ is a n-manifold.  

Although the definition of closed n-conifolds  is easiest to state, it is 
necessary to define a n-conifold with boundary to formulate generalized 
histories for the Hartle-Hawking initial state. Recall that the points  on the 
boundary of a n-manifold with boundary have neighborhoods that are
deformable to half of an n-ball. Thus
\proclaim Definition 2.3. A  n-dimensional conifold $X^n$, $n\ge 2$, is a 
metrizable space such that any point $x_0 \in X^n$ has either 1) a conical 
neighborhood or 2) an open neighborhood homeomorphic to an open subset of the 
half-space ${\bf R}_+^n$. \par
\noindent The boundary of a conifold is defined as the set of points which are 
mapped to boundary points of ${\bf R}_+^n$. It follows from this definition
that the boundary of a conifold $X^n$ is a closed (n-1)-manifold. Figures 
\ref{crp2} and \ref{rp3} are examples of conifolds with boundary.

Def.  2.3 is a purely topological definition of an n-conifold; it is important 
to realize that additional structure is needed to define differentiable 
functions and metrics on the conifold.  The definition, that of a smooth 
structure, needed to do so closely parallels that used in defining smooth 
manifolds. Recall that a smooth structure is given implicitly in the 
specification of an atlas, that is a collection of coordinate charts, on a 
manifold \cite{kn}. This specification is implicit as a given smooth structure
can be compatible with many different atlases. Additionally, there may be more
than one inequivalent smooth structure on a given topological manifold. A 
smooth structure on a conifold similarly is implicit in the specification of 
an appropriate atlas; intuitively, the difference lies in the treatment of the 
points at which it is not a manifold, that is at the singular set $S$. More 
precisely, given a  smooth structure on the manifold $X^n - S$ in an appropriate 
atlas,  one can induce a smooth structure on the corresponding conifold $X^n$ 
by an extension to the singular points. This construction is done in detail in 
\cite{I}. As for manifolds, a given topological conifold may have more than one
inequivalent smooth structure. For current purposes it suffices to observe that, 
as for manifolds, an explicit coordinatization of the n-conifold provides the 
necessary realization of the smooth structure needed for the definition of 
differentiable metrics and fields.
 
The set of smooth conifolds has many of the nice geometrical properties 
associated with smooth manifolds. In particular, smooth conifolds admit a 
natural notion of distance and  geodesics  represented in terms of a Riemannian 
metric $g$: 
\proclaim Definition 2.4. The Riemannian metric on a smooth conifold $X^n$ 
is  given by the metric $g$ on the manifold $X^n - S$ extended to the singular 
set $S$ by taking the cauchy completion of distance function induced by $g$ to 
these points.\par
\noindent This completion provides a well defined  geometry on $X^n$. 
Defining a scalar field on a conifold is easier as there is less structure 
involved:
\proclaim Definition 2.5. A scalar field $\phi$ on a conifold is function,
that is a map $\phi:X^n\to {\bf R}$.\par
\noindent One can also view the scalar field on a conifold as the extension of 
$\phi$ defined on the manifold $X^n - S$  to the singular points given by 
assigning a value to the map at these points.

The smooth structure on $X^n$  additionally allows one to define continuous and 
differentiable fields on conifolds in direct parallel to their definition on
smooth manifolds. Recall that a continuous function on a  manifold, is one 
for which the inverse image of open sets are open. Similarly, noting that 
an atlas defines the open sets  on the conifold, a function such as $\phi$ on 
a conifold will be continuous  if the inverse image of these open sets is open. 
A differentiable function on a manifold is defined with reference to its smooth 
structure as realized in a compatible atlas; $\phi$ is $C^k$ if its composition 
with the coordinate maps is at least $C^k$ everywhere on ${\bf R}^n$ \cite{kn}.
Clearly, one can extend this procedure to define differentiable functions
on a conifold $X^n$. Finally as each component of a metric is a function, one
can extend this procedure  to define continuous and differentiable metrics 
on $X^n$. 
 
Integration on a conifold $X^n$  is defined using the measure $d\mu(g)$
associated with metric $g$ at all manifold points and extending it to the 
singular points. As the singular set $S$ is a discrete countable set of points, 
the contribution of these points to the measure is zero.

Conifolds also admit a natural definition of curvature, again closely related
to its definition on manifolds. Recall that the Riemann curvature tensor on a 
smooth manifold with smooth metric can be defined in terms of parallel
transport of a vector around an infinitesimal closed curve. This construction 
of curvature can be immediately extended to define the curvature tensor and
its contractions at all points of $X^n - S$.  Finally, the limiting behavior of 
the curvature  onto each singular point in $S$ will result in a well defined 
extension of the curvature to this point. In particular, the scalar curvature 
$R$ on a conifold  is that of the metric $g$ on  the manifold $X^n - S$ 
extended to the singular points. 

More precisely, one carries out this extension using
a sequence of Riemannian manifolds $(M^n_k,g_k)$ that converge to the conifold
$(X^n,g)$ in the appropriate topology as discussed in section 6 of \cite{I}. 
This extension may lead to curvature singularities at the singular points; 
for example, \cite{fursol} uses a similar technique to derive the curvature 
singularity standardly represented as a delta function on a two dimensional 
cone. However,  observe that
the character of the curvature singularity  at singular points of conifolds
is dimension dependent. In particular one can show 
$\int_V Rd\mu (g)=\int_{V-p} R d\mu (g)$
for any volume $V$ inclosing a singular point $p\in S$ 
in  dimension greater than two \cite{curvature}. An intuitive
feel for this result
 follows by a dimensional argument on the sequence 
$(M^n_k,g_k)$ converging to $(X^n,g)$. In particular, suppose
$X^n$ has one singular point $p$ and take
 the $(M^n_k,g_k)$  to correspond to $(X^n,g)$ with the
 neighborhood of the singular point $p$ removed and 
capped with a space $C^n_k$ of  decreasing volume. Then as
$R$ has dimension  $l^{-2}$ and $V$ dimension $l^n$, the 
integral over the
cap, $\int_{C^n_k} R_kd\mu (g_k)$, has dimension $ l^{2-n}$. As the condition
that the volume vanishes  corresponds to $l\to 0$,  one expects the
integrated curvature to vanish on the cap for $n>2$. Though the full
derivation of this result is more involved, 
this procedure is easy to illustrate in a simple example as done 
in Appendix \ref{section:i}.
 Therefore such possible
curvature singularities do not contribute to integrated quantities in 
three or more dimensions.

Thus the Einstein 
action on a compact conifold $X^n$ with boundary $\Sigma^{n-1}$ for metric 
$g$ is 
\begin{equation}
I[g] =-  {1 \over 16\pi G} \int_{X^n} 
(R-2\Lambda) d\mu(g) -  {1 \over 8\pi G }\int_{\Sigma^{n-1}}  K d\mu(h)
\label{2.1}
\end{equation}
where $h$ is the induced metric, $K$  the extrinsic curvature and $d\mu(h)$ the
covariant volume element of the boundary. Similarly, the action of a  massive 
scalar field $\phi$ with arbitrary coupling to the scalar curvature is given by
\begin{equation}
I[g,\phi]
=\frac 12 \int_{X^n} (\nabla_a\phi \nabla^a \phi + (m^2 +\xi R)\phi^2 ) d\mu(g)
+\int_{\Sigma^{n-1}} \xi K\phi^2 d\mu(h).\label{2.2}
\end{equation}
Observe that this action contains a $\xi$ dependent boundary term. This term 
is needed for \ref{2.2} to generate the correct stress energy tensor for 
the coupled Einstein equations from the variations of the metric 
\cite{bandd1}.

From the above discussion, it is apparent that although conifolds are more 
general topological spaces than manifolds, they share many of the same 
properties.  In particular, conifolds admit a natural extension of the concepts 
of differentiability and curvature as needed for the definition of generalized 
histories for the Hartle-Hawking state.

\subsection{The Generalized Hartle-Hawking State}

The generalized Hartle-Hawking initial state for gravity coupled to positive 
cosmological constant and scalar field can now be formulated;
\begin{equation}
\Psi[\Sigma^{n-1}, h,\varphi] = \sum_{X^n}\int Dg 
\exp\biggl(-I[g]-I[g,\phi] \biggr) \label{2.3}
\end{equation} 
where $\Sigma^{n-1}$ is the closed boundary manifold, $h$ the induced metric 
and $\varphi$ the field value on the boundary. The boundary conditions for this 
state are specified in terms of conditions on the histories to be included in 
this sum; formally, these {\it generalized histories} consist of suitably 
regular, physically distinct metrics $g$ and field configurations $\phi$ on 
compact conifolds $X^n$ that have the correct induced values on the
boundary $\Sigma^{n-1}$.  

Of course, the monumental task in making any functional integral such as 
\ref{2.3} well defined is to make precise this vague specification of the 
space of histories. As discussed in \cite{I},  a more detailed specification
of the space of histories would be anticipated to include distributional 
metrics and fields. It is not clear at the present time what an appropriate 
set of distributional histories is for formulating amplitudes such as the 
generalized Hartle-Hawking state. Even so, the smooth structure and topology 
of the conifold $X^n$ will be carried in such distributional configurations;  
these configurations are generally defined as elements of an appropriate dual 
space of smooth regular test fields and metrics. As these smooth regular test 
fields and metrics are defined with respect to the smooth structure and 
topology of the conifold, this information will also be carried in the 
distributional metrics and field configurations. 

Fortunately for the purposes of this paper, a precise specification of
the space of histories is not necessary; all that is needed for a semiclassical 
evaluation of \ref{2.3} is the definition of  a regular classical history.
The notion of a regular classical geometry is already implicit in Def. 2.4 of 
a Riemannian metric on a conifold. It is also clear how to specify a regular
classical scalar field. However it is useful to further clarify 
these points.  

First, recall that a natural notion of a regular geometry on a closed manifold 
is given by geodesic completeness. A geodesically complete closed manifold is
one for which any  geodesic of finite parameter length can be extended to one 
with infinite parameter length \cite{kn}. This property implies that these is no 
physical pathology in the geometry as there are no points for which the 
geodesics terminate. Moreover, it is apparent that this regularity is 
independent of the coordinate charts on the manifold. Thus geodesic 
completeness provides a natural definition of a regular closed manifold. Now the 
definition of a geodesically complete closed conifold exactly parallels that of 
a geodesically complete closed manifold \cite{I}. It follows from the above 
discussion that geodesic completeness again provides a natural definition of a 
regular closed conifold. 

This is a nice characterization of regularity, but the case at hand does not 
involve closed conifolds but rather ones with boundary. Unfortunately, there 
are technical difficulties in defining a geodesically complete conifold
with boundary due to the fact that geodesics can terminate at the boundary 
without signalling any physical pathology. This is not surprising as exactly
the same problem is present in the case of manifolds with boundary. However, 
there is an equivalent  notion to geodesic completeness that can be directly 
extended to this case: metric completeness. A metrically complete conifold is 
one that is cauchy complete, that is any sequence of points has a convergent 
subsequence in the distance function induced by the metric. Moreover one can 
prove a generalization of the Hopf-Rinow theorem \cite{kn} to conifolds. This 
theorem establishes that a closed conifold is geodesically complete if and only 
if it is metrically complete \cite{I}. Therefore an equivalent definition of a 
regular closed conifold is one which is metrically complete. Moreover, the 
notion of metric completeness can be immediately applied to conifolds with 
boundary as limit points of any sequence that approaches the boundary are by 
definition included in the space. Now any Riemannian metric on a conifold is by 
Def. 2.4 metrically complete. Therefore Def. 2.4 already incorporates a natural 
notion of a regular conifold, metric completeness.

Note that metric completeness is a relatively weak condition as it allows the 
possibility of spaces with certain curvature singularities. For example, a disk 
with metric \hbox{$ds^2 = dr^2 + r^2 d\theta^2$} where $0\le r <r_0$,
$0\le\theta\le 2\pi - \alpha$ is a geodesically complete space that is singular 
at $r=0$. For this coordinatization, the curvature exhibits a delta function 
singularity at the apex of the cone. Now such a history  may very naturally be 
included in the space of histories as it corresponds to relatively nice
distributional history. Indeed, such a metric in two dimensions even exhibits 
finite action. However, as classical solutions of the Einstein equations, one 
might like to avoid inclusion of such histories. This can be done by placing 
additional restrictions on the regularity of the Riemann curvature or  its 
contractions. However, for the case of  positive cosmological constant such 
additional conditions are not needed; solutions to the Euclidean Einstein 
equations on closed conifolds are analytic \cite{I}. This analyticity provides 
the desired additional regularity of the curvature. Thus it suffices to define 
a regular conifold as one  with complete metric.

Given the above considerations, it is manifestly apparent that a suitable 
definition of a regular classical field is one which is continuous and bounded 
on $X^n$. Again this definition is not sufficient to restrict field 
configurations to ones that exhibit further desirable properties such as 
differentiability. However, again one anticipates that the solutions of the 
classical field equations  will exhibit the additional smoothness properties. 

At this point, it is clear how to state a sufficiently precise definition of a 
suitably regular classical history for semiclassical analysis: a  {\it regular 
classical generalized history} consists of a metrically complete metric $g$ 
and continuous, bounded scalar field $\phi$ on a smooth compact conifold $X^n$ 
that has the appropriate induced values $h$, $\varphi$ on the boundary 
$\Sigma^{n-1}$.

\subsection{Semiclassical Evaluation of the Generalized Hartle-Hawking 
Initial State}

As discussed by Hartle \cite{hartle}, a semiclassical evaluation of the 
Hartle-Hawking initial state for gravity coupled to a scalar field in the 
perturbative limit provides the connection between quantum cosmology and 
quantum field theory in curved space. Additionally it appears that the 
Hartle-Hawking boundary condition yields a particular choice of vacuum for 
the scalar field theory. A similar connection holds for the generalized 
Hartle-Hawking initial state formulated in terms of a sum over conifolds as 
demonstrated below. Furthermore, a particular virtue of the generalized
Hartle-Hawking state is that it produces semiclassical wavefunctions for
boundary manifolds that have no semiclassical solution for the original
Hartle-Hawking state formulated as a sum over manifolds.

A semiclassical evaluation of the generalized Hartle-Hawking state 
\ref{2.3} involves finding  an appropriately regular geometry and scalar 
field configuration corresponding to an extremum of the action on some compact 
conifold $X^n$. For actions of the form \ref{2.3} these extrema are 
solutions of the coupled Einstein equations 
\begin{mathletters}
\begin{eqnarray}
R_{ab}-\frac 12(R-2\Lambda) g_{ab} &=& 8\pi
G T_{ab} \label{2.4 a} \\ -\nabla^2 \phi - ( \xi R + m^2)\phi &=& 0 
\label{2.4 b}
\end{eqnarray}
\end{mathletters} 
where the stress energy tensor $T_{ab}$ is computed from variations of 
\ref{2.2} \cite{bandd1}. 

As these equations are coupled through the stress energy tensor, they are a set 
of highly nonlinear equations. Any  classical solution of them in the presence 
of inhomogeneous scalar field perturbations will in general lead to 
inhomogeneities in the  metric. Therefore these equations in general do not 
have a simple closed form solution. However, if one makes the additional 
approximation that the contribution to \ref{2.4 a} from the stress energy 
tensor can be neglected, these equations correspond to those for a quantum 
field theory on a fixed background Einstein conifold;
\begin{mathletters}
\label{2.5}
\begin{eqnarray}
R_{ab} &=&  \Lambda g_{ab}\label{2.5 a}\\ -\nabla^2 \phi
- \biggl( {4\xi\Lambda}  + m^2\biggr)\phi &=& 0.\label{2.5 b}
\end{eqnarray}
\end{mathletters}
This approximation corresponds to the physical situation where all components 
of the classical stress-energy tensor computed from the solution of \ref{2.5} 
are dominated by the cosmological constant at all times. This situation is a 
realistic approximation when the scalar field  boundary data $\varphi(\theta)$
can be treated as a small perturbation on a locally isotropic
background.\footnote{An alternate but equivalent view presented by Banks 
\cite{banks} is that one is computing the semiclassical evaluation of 
\ref{2.3} with $8\pi G$ as the small parameter. To lowest order in this 
parameter, \ref{2.5 a} is the equation describing the extremum of the action. 
To next order, one has the contribution of both fluctuations in the metric about 
this background and the contribution from the scalar field. The scalar field 
contribution to the semiclassical Hartle-Hawking initial state can also be 
evaluated in semiclassical approximation using \ref{2.5 b} as it first enters 
at this order.}

The metric of a classical history satisfying \ref{2.5 a} is typically Euclidean 
at small geometries and Lorentzian at large geometries. If there is more than 
one extremum of the action, the semiclassical approximation will consist of a 
superposition of the extrema although in practice often one keeps only the 
dominant contribution. 

It is difficult to further analyze the properties of the semiclassical
extrema without the context of a specific example. However, a typical
case that arises in such analysis is one for which the boundary metric
divides a closed Einstein conifold. For this case, the solution of
\ref{2.5 a} is analytic. Therefore choosing coordinates such that
$\tau = 0$ is the singular point, the metric near a conical singularity
$C(\Sigma^{n-1}_{0})$ takes the form
\[
ds^2=d\tau^2 +f(\tau) d\sigma^2
\]
where $f(\tau)\sim\tau^2 + O(\tau^3)$ and $d\sigma^2$ is a metric independent
of $\tau$ on the closed $\Sigma^{n-1}_{0}$ manifold. Now Cheeger 
\cite{cheeger} shows that the heat kernels of laplacians are well defined and 
computable for spaces that are cones over closed manifolds with metrics of 
this form. Moreover,  the space of differential forms carries the usual 
connections to the topology of the cone $C(\Sigma^{n-1}_{0})$. Clearly, a 
physical realization of such laplacians is given by free scalar field theory 
with action of form \ref{2.2}. Therefore, the regularity 
condition on the scalar field is sufficient in principle to yield a well 
defined semiclassical wavefunction in the Euclidean sector.

\section{The Semiclassical Initial State for 
 $RP^3$ Boundary with Round Metric and Scalar Field Perturbations}
\label{section:pm}

As demonstrated by Hartle \cite{hartle}, the original Hartle-Hawking initial 
state formulated as a sum over manifolds yields the Euclidean vacuum when 
evaluated semiclassically for conformally coupled massless scalar field 
perturbations on a de Sitter background. Laflamme \cite{laflamme} 
demonstrated that similar results hold for the minimally coupled scalar field.  
As shown below, the semiclassical approximation generalized Hartle-Hawking 
initial state yields a well behaved, unique vacuum state for $RP^3$ boundary 
with round metric and scalar field perturbations. Furthermore, this 
state cannot be obtained from the original Hartle-Hawking 
initial state as there is no spherically symmetric Einstein manifold with
round $RP^3$ boundary. This state is not identical to but has
a natural correspondence to the Bunch-Davies vacuum for a scalar field on 
$RP^3$ de Sitter spacetime for large geometries.  This derivation will be done 
in complete generality, that is for a massive scalar field  with arbitrary 
scalar curvature coupling.

The initial data for this wavefunction consists of the manifold $RP^3$ with its 
round metric of radius $a_0$ and inhomogeneous scalar field perturbation 
$\varphi(\theta)$. The round metric on $RP^3$ can be constructed from that on 
$S^3$ by an identification of antipodal points as illustrated in figure 
\ref{rpn}; a convenient coordinatization of this metric is
\begin{eqnarray}
 d\Omega^2&=&d\theta_{{\scriptscriptstyle 1}}^2+\sin^2\theta_{
{\scriptscriptstyle 1}}(
d\theta_{{\scriptscriptstyle 2}}^2+\sin^2\theta_{{\scriptscriptstyle 2}}
d\theta_{{\scriptscriptstyle 3}}^2)\nonumber\\
 0\le\theta_{{\scriptscriptstyle 1}}\le \pi ;
&\ & \ \ 0\le\theta_{{\scriptscriptstyle 2}}\le \pi  \ \ \ \ 
0\le\theta_{{\scriptscriptstyle 3}}\le \pi \label{3.1}
\end{eqnarray}  
It is clear the round metric on $RP^3$ is locally the same as that on $S^3$, 
differing only by having a reduced coordinate range. The boundary field data 
$\varphi(\theta)$ is implicitly assumed to be of small enough magnitude that 
the corresponding classical solution will have a sufficiently small stress
energy tensor for the approximation leading to \ref{2.5} to be valid. One 
first solves \ref{2.5 a} for the background geometry for all values of $a_0$. 
Then one solves \ref{2.5 b} using this background geometry for the scalar field.

It  follows from the explicit form of the metric \ref{3.1} that the 
manifold $RP^3$ has the same local properties as  $S^3$ with round metric. 
Therefore the Einstein equations for these two manifolds have the same local 
form and consequently the same local solution for locally identical initial 
data sets. In particular, for $a_0<1/H$ where $H^2 = \Lambda/3$, the equations 
of motion for an explicitly spherically symmetric metric of form 
\begin{equation}
ds^2 =  {1 \over H^2}d\tau^2 + a^2(\tau) d\Omega^2\ \ \ \ \ \label{3.2}
\end{equation} 
are 
\begin{equation}
{\partial_\tau^2 a\over a}
-\left({\partial_\tau a \over a}\right)^2 +  {1 \over H^2a^2} =0.\label{3.3}
\end{equation}
Thus a unique solution in a neighborhood of the initial $RP^3$ hypersurface is
\begin{equation}
a(\tau) =  {1\over H}
\sin\tau \label{3.4}
\end{equation} 
This solution has product topology $I\times RP^3$ for the range $0<\tau< \pi$. 
Extending the range of $\tau$ to $\tau=0$ results in the explicit 
identification of all points on the $RP^3$ spatial slice to one point.
Therefore this extension yields the compact conifold $C(RP^3)$ illustrated in 
Figure \ref{rp3}. 

It is easy to demonstrate that this conifold is not a manifold using properties 
of its fundamental group. First recall that the fundamental group of any 
4-manifold $M^4$ satisfies the relation $\pi_1(M^4-\{p\}) = \pi_1(M^4)$  for 
any point $p\in M^4$. This is due to the fact that in three or more dimensions, 
curves can always be moved around an isolated point without ever passing 
through the point itself. Now assume that $C(RP^3)$ is a manifold and take the 
point $p$ to be the apex of the cone. Note that by construction 
$C(RP^3) - \{p\} = I\times RP^3$. 
Hence $\pi_1(C(RP^3)-\{p\}) = \pi_1(I\times RP^3)) = Z_2$ as the 
fundamental group of $RP^3$ is $Z_2$. But by definition, $C(RP^3)$ is 
contractible which implies that $\pi_1(C(RP^3)) = 1$. Therefore, by 
contradiction, it follows that $C(RP^3)$ is not a manifold. Thus the solution 
of the Euclidean Einstein equations yields a nonmanifold stationary point of 
the action.

Note that uniqueness of the solution to  \ref{3.3} implies that
there is no spherically symmetric Einstein manifold with round
$RP^3$ boundary. In addition, as given in detail in Sec. 3 of\cite{I}, one
can use the fact that Einstein manifolds are analytic to argue that
there is no Einstein manifold with round $RP^3$ boundary. Therefore, the 
Einstein conifold constructed above is the only known stationary point of the 
action. Thus its contribution is essential to producing a semiclassical 
amplitude for the generalized Hartle-Hawking state. It is particularly so for 
the case at hand as it will be seen that a spherically symmetric background is 
necessary to obtain the Bunch-Davies vacuum.

In general, there are two possible positions for the $RP^3$  boundary with 
scale factor $a_0$ in the Euclidean solution \ref{3.2} that yield the
correct induced metric; one at coordinate position $\tau_0<\pi/2$  and one at
$\tau_0>\pi/2$. According to \cite{hh}, the one that dominates in 
semiclassical approximation is that with least action  
\begin{equation}
I(a_0)= - {\pi \over 4GH^2} [(1-H^2a_0^2)^\frac 32 -1] \label{3.5}
\end{equation}
corresponding to $\tau_0<\pi/2$. This action differs by a factor of $1/2$ from 
that of the familiar $S^3$ boundary case due to the difference in volume of a 
round $S^3$ and $RP^3$.

For $Ha_0>1$, there are no real Euclidean extrema.  Instead, there are
two complex extrema corresponding to the Lorentzian $RP^3$ de Sitter solution
\begin{equation}
ds^2 = - {1 \over H^2}d t^2 + a^2(t) d\Omega^2\label{3.6}
\end{equation}
where the scale factor is now
\begin{equation}
a(t) =  {1\over H} \cosh t.\label{3.7}
\end{equation}
This solution is a globally hyperbolic spacetime of topology 
${\bf R}\times RP^3$. It is clearly related to the usual globally
hyperbolic de Sitter spacetime of topology ${\bf R}\times S^3$, henceforth
also known as $S^3$ de Sitter spacetime; the $RP^3$ de Sitter
 is constructed from $S^3$ de Sitter  by identification 
of antipodal points on each spatial $S^3$. Both extrema have complex action 
\begin{eqnarray}
I(a_0) &=& \mp iS(a_0)\nonumber\\
S(a_0) &=&
{\pi \over 4GH^2}(H^2a_0^2-1)^{\frac 32}.\label{3.8}
\end{eqnarray}

A useful way to characterize the connection of these extrema to the  Euclidean 
de Sitter solution is to use the fact that this solution is an analytic function 
of $\tau$; any continuation of $\tau $ to complex value in the Euclidean 
solution \ref{3.4} yields a solution to the equation of motion  \ref{3.3} for 
the same complex value. But the Lorentzian Einstein equations result from 
precisely such a continuation of $\tau$. Therefore one can connect the Euclidean 
solution  to the Lorentzian one  by finding a contour of complex 
$\tau$  that extends one to the other and results in  real spatial geometry for 
all $\tau$. The appropriate contour is to take $\tau$ real for range 
$0\le\tau\le\pi/2$ and then to take $\tau$ complex along the path  
$\tau = \pi/2 \pm it$. The resulting complex spacetime is therefore Euclidean 
for $\tau < \pi/2$ and Lorentzian for $\tau > \pi/2$. Observe that the 
generalized Hartle-Hawking boundary condition is naturally enforced for the 
complex extrema by the regularity condition on the Euclidean sector of the 
complex spacetime. 

The two possible contours of $\tau$ in the construction of the complex spacetime 
correspond to the two Lorentzian extrema. That given by the continuation 
$\tau\to \pi/2 +it$ corresponds to the expanding phase of $RP^3$ de Sitter 
with action $I(a_0)= -iS(a_0)$ and that for $\tau \to \pi/2 - it$ corresponds 
to the contracting phase with action $I(a_0)= iS(a_0)$. This unified view of 
the extrema of the action will be especially useful for the computation of the 
scalar field modes.

\subsection{The Classical Scalar Field Solution}
\label{section:mode}

Given the determination of the background geometry, one can now solve
for the scalar field. On a background of form \ref{3.2}, the scalar field 
equation becomes
\begin{equation}
-\biggl[{H^2\over a^3(\tau)}\partial_\tau(a^3(\tau) \partial_\tau)
+{1\over a^2(\tau) }D^2  - (m^2 + 12\xi H^2)\biggr]\phi(\tau,\theta) = 
0.\label{3.9}
\end{equation}
To solve this equation for a regular scalar field  with boundary value
$\varphi(\theta)$, it is useful to first solve \ref{3.9} for a complete set of
regular modes.  It is natural to do so in terms of a complete set of
orthogonal functions on  $RP^3$. As discussed in Appendix \ref{section:h}, the 
set of hyperspherical harmonics $Q_{(n)}(\theta)$, with principle eigenvalue 
$n$ restricted to odd integers form such a complete set.

For $a_0<1/H$,  the background geometry is the $C(RP^3)$ Einstein conifold. 
For this parameter range, is useful to write the modes in the form
\begin{equation}
v^{\scriptscriptstyle E}_{(n)}(\tau,\theta) = 
{r_{n}(\tau)\over \sin(\tau)} Q_{(n)}(\theta).\label{3.10}
\end{equation}
If these are to be a solution of \ref{3.9} for $a(\tau)$ given by \ref{3.4}, 
then $r_n(\tau)$ must satisfy 
\[
\biggl[- {1\over \sin\tau} \partial_\tau(\sin\tau \partial_\tau) 
+ {n^2\over \sin^2\tau}  + ( {m^2\over H^2} + 12 \xi -2)\biggr]
r_{n}(\tau)
=0.\] %\label{3.11}
The general solution of this equation is a linear combination of the
Legendre polynomials $ P_{\nu-\frac 12}^{-n}(\cos \tau)$ and 
$ Q_{\nu-\frac 12}^{-n}(\cos \tau)$. The requirement that \ref{3.10} be 
regular over the background Einstein conifold fixes the coefficient of 
$ Q_{\nu-\frac 12}^{-n}(\cos \tau)$ to vanish as it diverges as $\tau \to 0$ 
for all $n$  \cite{as}. Thus 
\begin{equation}
r_n(\tau) =A_n P_{\nu-\frac 12}^{-n}(\cos \tau)
\label{3.12}
\end{equation}
where $\nu = \sqrt{\frac 94 - {m^2\over H^2} - 12\xi}$ and $A_n$ is a 
normalization constant.

For $a_0>1/H$, the background solution is no longer Euclidean, but is complex. 
The scalar field equation in the Lorentzian sector of the background solution 
is
\begin{equation}
\biggl[{H^2 \over a^3(t)}\partial_t(a^3(t)\partial_t)
- {1 \over a^2(t)} D^2  + (m^2 + 12\xi H^2)\biggr]\phi(t,\theta) = 
0\label{3.13}
\end{equation}
with $a(t)$ given by \ref{3.7}. 
To explicitly construct the modes in this sector it is conventional to change 
the variable $t$ in \ref{3.13} to  $\eta$ where $\sin \eta = 1/\cosh t$. 
Note that $t\to \infty$ corresponds to $\eta\to \pi$. The modes are then 
\begin{equation}
v^{\scriptscriptstyle L}_{(n)}(\eta,\theta) =
\sin^{\frac 32} \eta\ \rho_{n}(\eta)Q_{(n)}(\theta).\label{3.14}
\end{equation}
These modes will satisfy the  Lorentzian scalar field equation \ref{3.13} for 
$\rho_n(\eta)$ that satisfies
\[
\biggl[{ 1\over \sin \eta}\partial_\eta(\sin \eta\partial_\eta)  +
\frac 1{\sin^2\eta}( {m^2\over H^2} + 12\xi -\frac 94) +n^2 -
\frac 14\biggr]\rho_n(\eta) =0.\] %\label{3.15}
This equation has the familiar general solution \cite{c+t,bandd} 
\begin{equation}
\rho_n(\eta) = B_n( P^\nu_{n-\frac 12}(-\cos\eta) + \kappa 
Q^\nu_{n-\frac 12}(-\cos\eta))\label{3.16}
\end{equation}
where $B_n$ and $\kappa$ are arbitrary constants.

The remaining task is to find these constants as determined by the generalized 
Hartle-Hawking boundary condition. Formally, they are determined directly from
the analytic continuation of the regular Euclidean field solution \ref{3.10}
with $r_n(\tau)$ given by \ref{3.12}; as this analytic continuation directly 
yields the Lorentzian equation \ref{3.13}, it also must yield the regular mode 
solutions of form \ref{3.14}. Unfortunately it is difficult explicitly carry 
out this continuation to directly relate this solution to one of form 
\ref{3.16} except for the special case ($\xi=1/6$,  $m^2 = 0$) as the modes in 
the Euclidean and Lorentzian sectors are expressed in terms of different 
variables and different Legendre polynomials. However, one can determine the 
constants as set by this continuation by exploiting the uniqueness of solutions
to the Legendre equation. The  unique solution to \ref{3.13} is specified by 
giving $v^{\scriptscriptstyle L}_{(n)}(\eta,\theta)$ and its derivative at a 
specified value of $\eta$. In particular, one can give this data at the 
transition point  between the Euclidean and Lorentzian sectors. 
As the Lorentzian field equation \ref{3.13} arises as the analytic continuation 
of the Euclidean one, the correct data to specify at this point is set by the 
regular Euclidean solution \ref{3.10} and its derivative evaluated at the 
transition point. 

The exact relation between the Lorentzian and Euclidean data depends on the
complex background solution. For the case of the expanding $RP^3$ de Sitter
spacetime corresponding to the continuation $\tau = \pi/2 + it$  it is
\begin{mathletters}
\begin{eqnarray}
v^{\scriptscriptstyle L}_{(n)}(\pi/2,\theta)&=&
v^{\scriptscriptstyle E}_{(n)}(\pi/2,\theta) \label{3.17 a}\\
- i \sin\eta\partial_\eta v^{\scriptscriptstyle L}_{(n)}(\eta,\theta)
\biggl|_{\eta=\pi/2} &=& 
\partial_\tau v^{\scriptscriptstyle E}_{(n)}(\tau,\theta)\biggl|_{\tau=\pi/2}.
\label{3.17 b}
\end{eqnarray}
\end{mathletters}
The dependence on the complex background is carried in the change of variables
\hbox{$\partial_\tau =  - i \sin\eta\partial_\eta$} used to express the 
derivative on the left hand side of \ref{3.17 b} in terms of $\eta$.

The constant $\kappa$ can be determined by taking the ratio of the appropriate 
sides of  \ref{3.17 a} by those of \ref{3.17 b}: 
\[
 - {i
 \partial_{ \eta} (\sin^{\frac 32}\eta\rho_n(\eta)) 
 \over \sin^{\frac 12} \eta \rho_n(\eta)}
\biggl|_{\eta=\pi/2} = {\sin \tau\over r_n(\tau)}
 \partial_{ \tau} \biggl( {r_n(\tau)\over \sin \tau}\biggr)
\biggl|_{\tau=\pi/2}.
%\label{3.18}
\]
Evaluating in terms of the  explicit forms  \ref{3.12} and \ref{3.16} 
yields
\[
 -i {\partial_x P^\nu_{n-\frac 12}(x) +\kappa 
\partial_x Q^\nu_{n-\frac 12}(x)\over P^\nu_{n-\frac 12}(x) +\kappa 
 Q^\nu_{n-\frac 12}(x)}\biggl|_{x=0}
 =
  -{\partial_x P^{-n}_{\nu - \frac 12}(x)\over
P^{-n}_{\nu - \frac 12}(x)}\biggl|_{x=0}.\]
Each side of this expression can be evaluated using identities for Legendre 
functions and their derivatives at zero argument \cite{as}, producing 
\begin{eqnarray}
-2 i\ 
{\frac 2\pi \sin\pi\alpha_n +\kappa \cos \pi\alpha_n\over
\frac 2\pi \cos\pi\alpha_n -\kappa \sin \pi\alpha_n}
{\Gamma(\alpha_n+1)\over \Gamma(\alpha_n+\frac 12)}&&
{\Gamma(n-\alpha_n+\frac 12)\over
\Gamma(n-\alpha_n)\ \ }=\nonumber\\
&&-2\tan  \pi (\alpha_n - n) 
{\Gamma(\alpha_n+1 )\over \Gamma(\alpha_n+\frac 12)}
 { \Gamma(\alpha_n+1-n )\over
\Gamma(\alpha_n+\frac 12-n)}\nonumber %\label{3.19}
\end{eqnarray}
where \hbox{$\alpha_n = \frac n2 + \frac \nu 2 - \frac 14$}.
This expression can be simplified considerably by using the identity 
$\Gamma(z)\Gamma(1-z) =\pi\csc\pi z$ to reexpress one of the Gamma function 
ratios on the right hand side of this equation;
\[
{\Gamma(\alpha_n+1-n )\over \Gamma(\alpha_n+\frac 12-n)}=
 {\sin\pi(\alpha_n+\frac 12-n)\over
 \sin\pi(\alpha_n+1-n)}
{\Gamma(n-\alpha_n+\frac 12)\over\Gamma(n-\alpha_n )}.
\]
Using this, the previous equation simplifies to
\[
 -i\frac{\frac 2\pi \sin\pi\alpha +\kappa \cos \pi\alpha}
{\frac 2\pi \cos\pi\alpha -\kappa \sin \pi\alpha}=1 .
\]
The solution of this equation yields $\displaystyle \kappa =  \frac 2\pi i$. 
Therefore, for the expanding $RP^3$ de Sitter background corresponding to the 
continuation $\tau = \pi/2+it$ 
\begin{equation}
\rho_n(\eta)=B_n(P^v_{n-\frac 12}(-\cos\eta) +
\frac 2\pi i Q^\nu_{n-\frac 12}(-\cos\eta)).
\label{3.20}
\end{equation}
Observe that this mode takes the form of the complex conjugate of the mode 
solutions \ref{B2} for a scalar field in the Bunch-Davies vacuum on $RP^3$ 
de Sitter. It will be seen that this correspondence is precisely that needed to 
establish the connection between the generalized Hartle-Hawking state and the 
Bunch-Davies vacuum.

The case of  the contracting $RP^3$ de Sitter solution given by the analytic 
continuation $\tau = \pi/2 -it$ follows immediately from the above derivation. 
Observe that \ref{3.17 b} will differ by a sign as 
$\partial_\tau =  i \sin\eta\partial_\eta$ for this continuation. Now all 
further manipulations to derive $\kappa$ are algebraic; therefore it follows 
immediately that the modes for the contracting $RP^3$ de Sitter extremum are
of the form \ref{3.14}  with 
$\rho_n(\eta)=B_n(P^v_{n-\frac 12}(-\cos\eta) -
\frac 2\pi i Q^\nu_{n-\frac 12}(-\cos\eta))$. As the Legendre functions
are real for real coefficients, these modes can be conveniently written
$\rho^*_n(\eta)$, the complex conjugate of the modes for the expanding $RP^3$ 
de Sitter extremum.

At this point, the regular modes for the scalar field equation on
background extrema for all $a_0$  have been constructed. One can now find the  
regular classical solution  of the scalar field equation by expanding it in 
terms of these modes. First,  the boundary scalar field perturbation 
$\varphi(\theta)$ can be expanded in terms of the  hyperspherical harmonics
of odd principle eigenvalue $n$, a complete set of orthogonal
functions on $RP^3$; 
\[
\varphi(\theta)=\sum_{\{odd\  n\}}\varphi_{(n)} Q_{(n)}(\theta)
\]
where ${\{odd\  n\}}$ indicates a sum over all  elements in the irreducible 
representations of the odd $n$ hyperspherical harmonics. The coefficients in 
this expansion are related to the perturbation by 
\begin{equation}
\varphi_{(n)} =\int 
\varphi(\theta)Q_{(n)}(\theta)d\mu({c})\label{3.21}
\end{equation}
where $d\mu(c)$ is the volume element for the unit metric \ref{3.1} on 
the round $RP^3$. 

For boundary data with $a_0<1/H$, the regular classical solution is expanded 
in terms of the Euclidean modes
\begin{equation}
 \phi(\tau,\theta)=\sum_{\{odd\  n\}}q_{(n)}
 v^{\scriptscriptstyle E}_{(n)}(\tau,\theta).
 \label{3.22}
\end{equation}
The coefficients $q_{(n)}$ are related to the boundary data by requiring that 
\ref{3.22} equal \ref{3.21} at the $RP^3$ boundary; one finds 
\begin{equation}
q_{(n)} = \frac {Ha_0}{r_n(\tau_0)} \varphi_{(n)} \label{3.23}
\end{equation}
where $\tau_0 = \sin^{-1} Ha_0$.
Evaluating  the classical field action \ref{2.2} for this solution 
yields
\begin{equation}
I_{cl}(a_0,\varphi)= \frac 12 {Ha_0^3} \sum_{\{odd\  n\}} 
\biggl[\frac {\partial_\tau r_n(\tau)} {r_n(\tau)}+(6\xi-1)
\frac {\cos\tau}{\sin\tau}\biggr]_{\tau=\tau_0} \varphi_{(n)}^2.\label{3.24}
\end{equation}

For boundary data with $a_0>1/H$, the regular classical solution is found in 
terms of the Lorentzian modes. For the expanding $RP^3$ de Sitter background 
given by the continuation $\tau = \pi/2 + it$
\begin{equation}
 \phi(\eta,\theta)=\sum_{\{odd\  n\}}q_{(n)}
 v^{\scriptscriptstyle L}_{(n)}(\eta,\theta)
 \label{3.25}
\end{equation}
with the $\rho_n(\eta)$ in $v^{\scriptscriptstyle L}_{(n)}(\eta,\theta) $ is 
given by \ref{3.20}. The coefficients $q_{(n)}$ are again related to the 
boundary data; 
\begin{equation}
q_{(n)} = \frac 1{\sin^{\frac 32}\eta_0\rho_{n}(\eta_0)}{\varphi_{(n)}}
\label{3.26}
\end{equation}
where $\eta_0 = \sin^{-1}1/({Ha_0})+\pi/2$. Observe that the $\pi/2$ term  
enters this relation as $\eta_0>\pi/2$  by its definition.

The classical action for this solution again is  obtained by evaluating
\ref{2.2}; 
\begin{eqnarray}
I_{cl}(a_0,\varphi) &=& -iS_{cl}(a_0,\varphi)\nonumber\\
S_{cl}(a_0,\varphi)&=& \frac 12 {Ha_0^3} \sum_{\{odd\  n\}} 
\biggl[ {\partial_\eta (\sin^{\frac 12}\eta\rho_n(\eta))\over 
\sin^{\frac 12}\eta\rho_n(\eta)}-(6\xi-1)
{\cos\eta}\biggr]_{\eta=\eta_0} \varphi_{(n)}^2.\label{3.27}
\end{eqnarray}

The solution for the contracting $RP^3$ de Sitter background given by the
continuation $\tau = \pi/2 -it$  is constructed similarly. First, the modes 
$\rho_n(\eta)$ in the expansion of the scalar field \ref{3.25} and in 
the expression for the coefficients $q_{(n)}$ are replaced by $\rho^*_n(\eta)$. 
It is  clear that as the boundary data is real, the solution for this case is 
given by the complex conjugate of \ref{3.25}. Additionally, the sign of 
the classical action evaluated this extremum is opposite to that for expanding 
$RP^3$ de Sitter background. Thus 
\begin{equation}
I_{cl}(a_0,\varphi) = iS^*_{cl}(a_0,\varphi)\label{3.28}
\end{equation}
for the contracting $RP^3$ de Sitter background.

\subsection{The Semiclassical Wavefunction for Perturbative Fields}
 
The semiclassical evaluation of the generalized Hartle-Hawking state for round 
$RP^3$ boundary  and scalar field perturbations now follows directly. For 
$a_0< 1/H$, the semiclassical wavefunction to lowest order consists of the 
exponential of both the Euclidean gravitational action \ref{3.5} and 
Euclidean scalar field action \ref{3.24}
\begin{equation}
\Psi_{\scriptscriptstyle E}[RP^3; a_0,\varphi] 
\sim \exp(-I(a_0)-I_{cl}(a_0,\phi)).\label{3.29}
\end{equation}

The semiclassical wavefunction for $a_0> 1/H$ consists of the superposition of 
the contributions from both complex extrema. First, the contribution from the 
expanding $RP^3$ de Sitter solution is given by the exponential of the sum of 
Lorentzian gravitational action \ref{3.8}  and  scalar field action 
\ref{3.27}
\begin{eqnarray}
\Psi^{expanding}[RP^3; a_0,\varphi]&&\sim\exp(iS(a_0)+iS_{cl}(a_0,\phi))
\nonumber\\
&&\sim\exp(iS(a_0)){\cal Q}[RP^3;a_0,\varphi]\label{3.30}
\end{eqnarray}
 where 
\begin{equation}
{\cal Q}[RP^3;a_0,\varphi]\sim \exp(iS_{cl}(a_0,\phi))\label{3.31}
\end{equation}
is the semiclassical wavefunction for the scalar field contribution on the 
$RP^3$ de Sitter background. As seen in Appendix \ref{section:bd}, ${\cal Q}
[RP^3;a_0,\varphi]$  is the Lorentzian Bunch-Davies vacuum state in field 
representation. 

The contribution from the contracting solution is given by the exponential of 
the sum of the Lorentzian gravitational action \ref{3.8} and scalar field 
action \ref{3.28},
\begin{eqnarray}
\Psi^{contracting}[RP^3; a_0,\varphi]&&\sim\exp(-iS(a_0)-iS^*_{cl}(a_0,\phi))
\nonumber\\
&&\sim\exp(-iS(a_0)){\cal Q}^*[RP^3;a_0,\varphi].\label{3.32}
\end{eqnarray}
The manifest reality of the functional integral for the generalized 
Hartle-Hawking initial state implies that the superposition of these two 
contributions must result in a real wavefunction. Thus the semiclassical 
wavefunction for  $a_0>1/H$ to lowest order is
\begin{equation}
\Psi_{\scriptscriptstyle L}[RP^3; a_0,\varphi] 
\sim \exp( i\alpha+iS(a_0)){\cal Q}[RP^3;a_0,\varphi] +
\exp (-i\alpha -iS(a_0)){\cal Q}^*[RP^3;a_0,\varphi]\label{3.33}
\end{equation}
where $\alpha$ is a phase set by further analysis of the steepest descent 
contour used in the semiclassical approximation. This phase will not be
calculated here as it is not needed for the objectives of this paper.

This semiclassical wavefunction takes a particularly simple form for the case 
of the conformally coupled massless scalar field, ($\xi= 1/6$, $m^2 = 0$).
In the Euclidean sector the modes for the conformally coupled case \ref{3.12} 
are
\[
r_{n}(\tau) ={A_n\over \Gamma(1+n)}
\biggl( {{1-\cos\tau}\over {1+\cos\tau}}\biggr)^{\frac n2}.
\]
The modes for the Lorentzian sector can be constructed directly from 
the Euclidean modes by analytic continuation or equivalently by evaluation of 
\ref{3.16}
\[
\rho_n(\eta) = {B_n \exp {in\eta}\over \Gamma(1+n)
(\sin\eta)^{\frac 12}} 
\]
The evaluation of the classical action for the solution with 
boundary data \ref{3.21} can be carried out explicitly; one finds that it is 
the same for both the Euclidean and Lorentzian background,
\begin{equation}
I_{cl}(a_0,\varphi) = \frac 12 a_0^2\sum_{\{odd\  n\}} n
\varphi_{(n)}^2.\label{3.34}
\end{equation}
This action is manifestly real; therefore 
\begin{eqnarray}
\Psi_{\scriptscriptstyle E}[RP^3; a_0,\varphi] 
&\sim& \exp(-I(a_0)){\cal Q}[RP^3; a_0,\varphi]
\ \ \ \ \ \ \ \ \ a_0<1/H\nonumber \\
\Psi_{\scriptscriptstyle L}[RP^3; a_0,\varphi] 
&\sim& \cos(S(a_0)+\alpha){\cal Q}[RP^3; a_0,\varphi]
\ \ \ \ \ \ \ a_0>1/H\label{3.35}
\end{eqnarray}
where the scalar field wavefunction is the same in both the Euclidean
and Lorentzian sectors,
\begin{equation}
{\cal Q}[RP^3; a_0,\varphi]\sim
\exp(-\frac 12 a_0^2\sum_{\{ odd\  n\} }n\varphi_{(n)}^2 ).\label{3.36}
\end{equation}
It is easily shown that this state corresponds to the conformal vacuum which 
is equivalent to the Bunch-Davies vacuum for the massless conformally coupled 
scalar field.\footnote{Note that the form of \ref{3.35} differs from that 
found in \cite{hh} and \cite{hartle} by a factor of $a_0^2$; these papers  
rescale the scalar field by a factor of $a(\tau)$ to a form especially 
adapted to the conformally coupled case. }

\section{Discussion}
\label{section:disc}

The semiclassical approximation to the generalized Hartle-Hawking state for 
scalar field perturbations and round $RP^3$ boundary has several interesting 
properties. Although some of these properties are particular to this particular
state, others can be expected to hold for more general boundary configurations.

First it is interesting to note the importance of the surface term in the 
scalar field action \ref{2.2}. The need for this term is already clear 
from the requirement that the scalar field action reproduce the correct stress 
energy tensor. However the explicit calculation of the semiclassical state 
demonstrates that it is essential for the  reproduction of the conformal 
vacuum \ref{3.35}. Without this term, there would be an additional 
contribution to the classical action for the conformal field \ref{3.34}; this 
can be easily seen by evaluating \ref{3.27} for the conformal perturbations  
with $\xi$ set to zero. This extra term clearly does not result in the correct 
form for the conformally coupled massless scalar field wavefunction. Therefore, 
the explicit calculation of the conformal vacuum from the general form  
\ref{3.31} derived in this paper emphasizes the importance of including 
the correct surface term in the scalar field action \ref{2.2}. 
Furthermore, the presence of this term will be reflected in the form of
the Wheeler de Witt equation for gravity coupled to a scalar field
with arbitrary mass and scalar field coupling. The derivation of the 
Hamiltonian follows by standard methods from the computation of the canonical
momenta from the scalar field action  \ref{B6}. This, combined with
the canonical commutation relations, will result in $\xi$ dependent terms in
the scalar field portion of the Wheeler de Witt equation.

More significantly, for general coupling, the semiclassical generalized 
Hartle-Hawking state \ref{3.33} is not of product form in the Lorentzian 
sector; the scalar field extrema is different for expanding and contracting 
$RP^3$ de Sitter backgrounds. Moreover, the contribution \ref{3.30}for an 
expanding $RP^3$ de Sitter background spacetime on its own corresponds to the 
Bunch-Davies vacuum.  So what does the contribution \ref{3.32} on the 
contracting $RP^3$ background correspond to? Clearly, this state is the 
complex conjugate of the expanding state.  Now the  expansions of 
$P^\nu_{n-\frac 12}(-\cos \eta)$ and $Q^\nu_{n-\frac 12}(-\cos \eta)$
in terms of elementary trigonometric functions \cite{as} can be used to show 
that
\begin{eqnarray}
P^\nu_{n-\frac 12}(-\cos \eta)- {2 \over \pi}i
&&Q^\nu_{n-\frac 12}(-\cos \eta)=\nonumber\\
&&-i A^\nu_n (-\sin\eta)^\nu 
\sum_{k=0}^\infty B^\nu_{nk}
\exp( i(\nu + n + \frac 12 + 2k)(\eta+\pi))\nonumber
\end{eqnarray}
where $\displaystyle A^\nu_n={2^{\nu+1}\Gamma(\nu+n+\frac 12) 
\over \sqrt{\pi}\Gamma(n+1)}$ and $\displaystyle B^\nu_{nk}={(\nu+\frac 12)_k
(\nu+n+\frac 12)_k\over k!(n+1)_k }$ are constants independent of $\eta$.
Given this expression, it is  easy to demonstrate that
\[
\rho^*_n(\eta) = \hbox{\rm e}^{i\beta_n}\rho_n(-\eta)
\]
where $\beta_n$ is a real constant. From this result, it is clear that the 
modes on the contracting $RP^3$ de Sitter are equivalent to up to a phase
the time reversed Bunch-Davies modes on the expanding background. Therefore, 
the contracting state \ref{3.32} corresponds to the time reversed Bunch-Davies 
vacuum state. Thus the semiclassical wavefunction \ref{3.33} is a 
superposition of an expanding $RP^3$ de Sitter spacetime with scalar field 
perturbations in the Bunch-Davies vacuum and a contracting $RP^3$ de Sitter 
spacetime with perturbations in the time reversed Bunch-Davies vacuum. This 
form suggests that a natural correlation exists between expansion and the
Bunch-Davies vacuum. A measurement of a conditional probability relating this 
expansion to the scalar field state could be anticipated to produce the result 
that the generalized Hartle-Hawking state implies that expanding round $RP^3$
universes have fields in the Bunch-Davies vacuum.
 
Additionally,  observe that the semiclassical extremum for the scalar field 
is real for $a_0<1/H$ but is complex for $a_0>1/H$. This follows directly from 
the fact that the Lorentzian modes \ref{3.20} are complex. This may be 
something of a shock as the field configurations summed over in the generalized
Hartle-Hawking state are real. However, on reflection, this property of the
extrema is simply that expected from the deformation of the geometry to complex 
values used to construct the stationary points of the action. This is emphasized 
by the fact that these complex paths reproduce the Bunch-Davies vacuum state in 
field representation for the expanding $RP^3$ de Sitter background. It is still 
interesting to note that although this deformation results in real 3-geometries 
at all points along the complex path, it produces complex field configurations  
for large 3-geometries along the same path. This result may have interesting  
implications when viewed in the context of the difficult issue of conformal 
rotation needed to make Euclidean integrals for gravity such as the 
Hartle-Hawking state bounded below \cite{conformalguys}.

It is also important to note the limitations inherent in this 
semiclassical wavefunction. In order to carry out the computation, it was 
assumed that the boundary data for the scalar field $\varphi(\theta)$ was a 
small perturbation, that is small enough that its stress energy tensor could be 
ignored in the calculation of the gravitational background. This condition 
clearly implies that the semiclassical wavefunctions of form \ref{3.29} in 
the Euclidean sector and \ref{3.33} in the Lorentzian sector are not valid 
for all ranges of scalar field boundary data. Moreover, these semiclassical 
wavefunctions cannot be realistically used as an initial condition for solving 
the Wheeler de Witt equation to find the solution for all ranges of scalar 
field boundary data on round metrics as one anticipates that large scalar 
field fluctuations will generally cause inhomogeneities in the background 
geometry. Therefore, solving the Wheeler-de Witt equation for the restrictive 
case of locally spherically symmetric geometries and general inhomogeneous 
scalar field would be expected to produce rather artificial behavior in the 
resulting wavefunction.
 
Finally, the  computation in this paper can be directly generalized to 
establish a similar result for any space $S^3/\Gamma$ with round metric
where $\Gamma$ is a finite group. For example, the Poincar\'e sphere is
such a space where $\Gamma$ is the binary icosahedral group. The
arguments for a local solution of the Einstein equations immediately 
generalize to show the existence of a Einstein conifold of the form 
$C(S^3/\Gamma)$. One can then find a complete set of harmonics for 
$S^3/\Gamma$ by constructing homogeneous polynomials of $S^3$ that are 
invariant under the action of the group $\Gamma$. The extrema of the field 
equation can then be constructed in terms of these harmonics.
Furthermore, one  anticipates that the same procedure can be carried
out to establish similar results for higher spin fields on these spaces.

\section{Conclusions}

Although conifolds are a set of topological spaces more general than 
manifolds, it is clear that they admit natural notions of regularity 
for both metrics and scalar fields. In addition, the implementation of these 
notions of regularity lead to well defined semiclassical approximations
to the generalized Hartle-Hawking state. These semiclassical 
states are produced for boundary configurations that have no such semiclassical
wavefunctions in the original Hartle-Hawking state formulated as a sum over 
manifolds. In particular, the semiclassical evaluation of the generalized 
Hartle-Hawking state for a perturbative scalar field on round $RP^3$ 
boundary yields a state that is not identical to but directly corresponds to 
the Bunch-Davies vacuum for $RP^3$ de Sitter. The production of this state 
depends essentially on the existence of the spherically symmetric Einstein 
conifold and its inclusion in the sum over histories.

\acknowledgments

This work has been supported in part by NSERC.

\appendix
\section{The Scalar Curvature of $C(S^{n-1})$ with Conical Metric}
\label{section:i}

The  general derivation of the curvature at a singular point of
a conifold $X^n$ with metric $g$ 
using a sequence $(M^n_k,g_k)$ of metrics on Riemannian manifolds 
will be given in detail in a elsewhere \cite{curvature}. 
However, it is useful to motivate
$\int_{V} Rd\mu(g) = \int_{V-p} Rd\mu(g)$ for any volume $V$ enclosing
a singular point $p$ of the conifold $X^n$ for $n>2$ 
using a simple example. This example is that of the
cone over $S^{n-1}$ with the metric
\begin{equation}
ds^2=dr^2+q^2r^2 d\Omega^2_{n-1}\label{cmetric}
\end{equation}
where $d\Omega^2_{n-1}=c_{ij}d\theta^i d\theta^j$ is the round 
metric of unit radius on
$S^{n-1}$. For $q=1$ this metric is flat; it exhibits a singularity
at $r=0$ for other $q$. In two dimensions, $q$ is related to
the deficit angle $\alpha$ by $q=1-\alpha/2\pi$. We will
consider only the case $q<1$ though it will
be clear how to generalize the results to $q>1$.
Clearly this example is one of a manifold with
a continuous metric not differentiable at $r=0$ rather than
that of a conifold with a topologically singular point.
However, the method used to compute the curvature using the
limit is the same clearly carries over to the more general
conifold case.

One can define a sequence of differentiable
metrics on $C(S^{n-1})$that converge to the metric (\ref{cmetric})
by cutting off the apex of the
cone at some radius $a$ and gluing on a cap. For $q<1$ a natural
choice for the cap consists
of a portion of an n-sphere with round metric
\begin{equation}
ds^2=R^2(d\xi^2+\sin^2\xi d\Omega^2_{n-1}).\label{round}
\end{equation}
 One matches the cap to the cut-off cone in 
a continuous and differentiable way by equating both the
metric and the extrinsic
curvature of their boundaries; this condition
will determine $R$ in terms of the other quantities.
One then takes $a\to 0$ to find the limiting values of the scalar
curvature and related quantities.

Explicitly, define a sequence of  metrics
$g_a$ given by an n-sphere with boundary (n-1)-sphere of radius $a$
with round metric (\ref{round}) and a cone with boundary (n-1)-sphere of
radius $a$ with metric (\ref{cmetric}). The boundary (n-1)-sphere of
the n-sphere 
 occurs
at coordinate $\xi_0 = \sin^{-1}( a/R)$ and has induced metric 
$R^2\sin^2\xi_0 c_{ij}$ and
 extrinsic curvature
$K_{ij}= R \sin\xi_0\cos\xi_0 c_{ij}$. The corresponding boundary (n-1)-sphere
on the cone occurs at  $r_0=a/q$ with induced metric $q^2r^2_0 c_{ij}$  and
 extrinsic curvature
$K_{ij}= q^2 r_0  c_{ij}$.
A continuous match of these induced metrics 
 and extrinsic curvatures 
on these boundaries requires $R$ such that
\begin{eqnarray}
a&=&R\sin\xi_0\nonumber\cr
{\frac  {\sin \xi_0}{R\cos \xi_0}}&=& {\frac 1{r_0}}={\frac qa}\label{mc}
\end{eqnarray}
 which yields
\begin{equation}
R=\frac a{\sqrt{1-q^2}}\ .
\end{equation}
Thus the scalar curvature on the n-sphere cap is 
$R=(n-1)(1-q^2)/a^2$ and diverges as
$a\to 0$. The integral of the curvature over the cap is
\begin{eqnarray}
\int_VRd\mu(g)& = &\frac {(n-1)}{R^2}R^nI_q \nonumber \cr
I_q &=&\frac {2\pi^{\frac {n-1}2}}{\Gamma(\frac {n-1}2)}
\int_0^{\sin^{-1}\sqrt{1-q^2}}\sin^{n-1}\xi d\xi.
\end{eqnarray}
As $I_q$ is independent of $a$ one finds that 
\begin{equation}
\int_{V_a} Rd\mu(g) =  (n-1)(1-q^2)^{\frac {2-n}2}a^{n-2}I_q
\end{equation}
which vanishes as  $a\to 0$ for $n>2$. Therefore, in this limit, the
curvature at the singular point does not contribute to integrals over the volume, 
namely
\begin{equation}
\int_{V} Rd\mu(g) =\int_{V-p} Rd\mu(g) 
\end{equation}
for any volume $V$ encompassing the apex $p$ of the cone for $n>2$.
For $n=2$, one can easily compute that $I_q = 2\pi(1-q)$ and
\begin{equation}
\int_{V_a} Rd\mu(g)=2\pi(1-q) = \alpha
\end{equation}
as expected. Thus in two dimensions, the integral of the curvature over
the cap is finite in the limit $a\to 0$.

Note that if $q>1$, the matching conditions (\ref{mc}) fail to have a real
solution. This is not surprising as this situation
corresponds to a negative curvature singularity at the point $r=0$. The
 choice of a cap with constant positive scalar curvature is inappropriate
as it cannot be matched in a continuous and differentiable way to the cut off
cone.
It is clear that the appropriate choice for this case is a sequence of caps
with  metrics with constant negative 
scalar curvature. It is easy to see that with this choice,
  one again derives that the integrated
scalar curvature vanishes for $n>2$ and recovers the usual contribution in
two dimensions.

Finally, it is clear that 
\begin{equation}
\int_{V} fRd\mu(g) =\int_{V-p} fRd\mu(g) 
\end{equation}
for any function $f$ continuous on $V$ for dimension $n>2$ as well. 
Therefore this limiting procedure recovers the standard result for the two
dimensional cone while showing that the integral of the curvature singularity
in higher dimensions vanishes.

\section{Orthogonal Functions on $RP^3$}
\label{section:h}

A complete set of orthogonal functions can be readily constructed on $RP^3$. 
These functions are a subset of the hyperspherical harmonics, a complete set of 
orthogonal functions on $S^3$. To construct the $RP^3$ hyperspherical harmonics, 
one uses the fact that $RP^3$ with its round metric can be constructed not only 
topologically but geometrically from $S^3$ with its round metric by 
identification of antipodal points.

First recall that one can geometrically embed $S^3$ with its round metric in
four dimensional Euclidean flat space as the surface $r^2 = \delta_{ab} x^ax^b$
where $x^a$ are the usual cartesian coordinates and $r$ is a real constant. 
Then $RP^3$ with its round metric is explicitly realized via the identification 
$x^a \to -x^a$, on this embedding  as illustrated in figure \ref{rpn}. This 
identification is precisely the parity transformation on the coordinates.

A complete set of orthogonal functions, the hyperspherical harmonics, are 
induced on $S^3$ by the set of homogeneous polynomials \cite{lk}, 
\begin{equation}
Q_{(n)}(\theta)={1\over r^{n-1}}A_{abcd\ldots}x^ax^bx^cx^d\ldots
\label{A1}
\end{equation}
where $A$ is a real symmetric constant traceless tensor with $n-1$ indices. 
They are irreducible infinite dimensional representations of $O(4)$, the
rotation group of $S^3$. The notation $(n)$ is of abbreviated form; it
indicates both the order $n$ and the other eigenvalues of the casimir operators 
labeling the orthogonal elements of this subspace of $O(4)$. One can readily 
show from \ref{A1} that the hyperspherical harmonics are eigenfunctions of
the Laplacian on $S^3$ with its round metric; explicitly
\begin{equation} 
D^2 Q_{(n)}(\theta)=-(n^2-1)Q_{(n)} (\theta)
\end{equation}
for unit radius. The dimension of the nth irreducible representation is $n^2$. 

Now to construct the $RP^3$ hyperspherical harmonics, observe that by 
construction, the $Q_{(n)}(\theta)$ are even under the identification 
$x^a \to -  x^a$ for odd values of $n$. These even hyperspherical harmonics
will be smooth, single valued functions on $ RP^3$ by virtue of the fact that 
this identification produces the manifold with its round metric. This is true 
for all elements of each subspace of odd order $n$ as no other restrictions on 
the tensor $A$ in \ref{A1} are imposed by the identification. Consequently 
the hyperspherical harmonics $Q_{(n)}(\theta)$ for odd $n$ are a complete set 
of states for expanding functions on $RP^3$. 

\section{The field representation of the Bunch-Davies vacuum for $RP^3$
de Sitter spacetime} 
\label{section:bd}

It is well known that is no preferred vacuum for a scalar field theory on
 de Sitter spacetime. Instead there is a one complex parameter family of 
de Sitter invariant vacuum states \cite{c+t} for the theory. One of these, the
Bunch-Davies vacuum, is a particularly natural choice. It is constructed in 
Lorentzian scalar field theory by requiring that the high momentum modes at 
late times approach their corresponding flat space form. A scalar field on
$RP^3$ de Sitter spacetime exhibits similar properties; there is no timelike 
killing vector on this spacetime  and consequently there is no preferred vacuum. 
Again, the Bunch-Davies vacuum can be constructed for this spacetime following 
the same method used in the $S^3$ de Sitter case. This construction of the 
Bunch-Davies vacuum for $RP^3$ de Sitter spacetime will be carried out in Fock 
space representation. Then the Fock space representation will be converted to 
field representation for comparison to results from the generalized 
Hartle-Hawking state.

In Fock space representation, the scalar field is an operator acting on the 
Hilbert space of field configurations,
\begin{equation}
\phi(t,\theta) = \sum_{\{odd \ n\}}b_{(n)}u_{(n)}(t,\theta) +b^\dagger
_{(n)}u^*_{(n)}(t,\theta)\label{B1}
\end{equation}
where $u_{(n)}$ is a mode solution of the Lorentzian field equation \ref{3.13} 
on $RP^3$ de Sitter corresponding to the Bunch-Davies vacuum. The $RP^3$ 
topology is carried by the restriction to odd  $n$ hyperspherical harmonics in
the expansion of the modes. The particle creation and annihilation operators 
satisfy the commutation relations
\[
\bigl[ b_{(n)},b^\dagger_{(n')} \bigr]=\delta_{(n)(n')}
\ \ \ \ \ \ \ \ 
\bigl[ b_{(n)},b_{(n')}\bigr] =
\bigl[b^\dagger_{(n)},b^\dagger_{(n')}\bigr]=0
\]
where $\delta_{(n)(n')}$ is a product of Kronecker delta functions that 
vanishes unless both sets of eigenvalues  are identical. The modes 
$u_{(n)}(t,\theta)$ are expanded
\begin{equation}
u_{(n)}(t,\theta)=U_n(t)Q_{(n)}(\theta)\label{B2}
\end{equation}
where the explicit form of $U(t)$ for the Bunch-Davies vacuum is most easily 
given in terms of the variable $\eta$ defined in Section \ref{section:mode},
\begin{equation}
U_n(\eta)=B_n \sin^{\frac 32}\eta
\biggl(P^v_{n-\frac 12}(-\cos\eta) -\frac 2\pi i Q^\nu_{n-\frac 12}
(-\cos\eta)\biggr).\label{tpart}
\end{equation}
Observe that $U_n(\eta)= \sin^{\frac 32}\eta \rho^*(\eta)$ terms of the modes 
\ref{3.20} of Section \ref{section:mode}. The modes \ref{B2} are 
orthonormal in the inner product
\begin{equation}
\int H a^3(t)  \biggl(u^*_{(n)}(t,\theta)\partial_t u_{(n')}
(t,\theta) -u_{(n')} (t,\theta) \partial_t u^*_{(n)}
(t,\theta)\biggr)d\mu(c)
=i\delta_{(n)(n')}.\label{B3}
\end{equation}

The vacuum is defined as the state containing no particles; it is thus 
annihilated by $b_{(n)}$ for all values $(n)$. In Fock space representation 
this is expressed as
\begin{equation}
b_{(n)}|0> = 0.
\label{B4}
\end{equation} 
Although this property of the vacuum might appear purely formal, it is  most 
definitely not. This relation can be used in conjunction with the definition
of the scalar field operator \ref{B1} and the commutation relation to compute 
quantities such as the two point function that are related to physically 
measurable properties of the vacuum state \cite{bandd}. Therefore, the 
specification of the modes and property \ref{B4} suffice to completely 
determine the Bunch-Davies vacuum for $RP^3$ de Sitter spacetime. Moreover, 
this defining property of the vacuum state  can be used to solve for the 
vacuum state in field representation.
 
In field representation the states are functionals of the field configuration
$\varphi(\theta) $ on a given spatial hypersurface at time $t$ where $t$ is the 
position of the $RP^3$ spatial hypersurface in the classical Lorentzian
background solution \ref{3.6}. The operators on these states are expressed in 
terms of the field configuration and functional derivatives with respect to 
the field configuration. Thus one can find the Bunch-Davies vacuum in field 
representation by reexpressing \ref{B4} in field representation and solving the
resulting functional equation. To do so, one must find the field representation 
of the annihilation operator $b_{(n)}$.
  
To do so, one needs to find $\pi(t,\theta)$, the momentum conjugate to the 
scalar field $\phi(t,\theta)$ in terms of the creation and annihilation
operators. One does so from its classical form. The classical Lorentzian 
scalar field action for $RP^3$ de Sitter background metric \ref{3.6} is
\begin{eqnarray}
S(a,\phi)
=\frac H2 \int  a^3\biggl(
(\partial_t\phi)^2 -{(D\phi)^2 \over H^2 a^2} - ({m^2 \over H^2}
+12 \xi)\phi^2 \biggr)
&& dtd\mu(c)\nonumber\\
&&+3H\xi\int_{RP^3} a^2 \partial_t a \phi^2 d\mu(c).  \label{B5}
\end{eqnarray}
The surface term can be expressed as a volume term by differentiation, 
resulting in the action in terms of a lagrangian density
\begin{eqnarray}
S(a,\phi)=&&\int  {\cal L }dt d\mu(c)\nonumber\\
{\cal L}=&&\frac H2 a^3\biggl(
(\partial_t\phi)^2+ 12 \xi \phi\partial_t\phi{\partial_t a\over a}
 -{(D\phi)^2\over H^2a^2} \nonumber \\ &&\ \ \ \ \ \ \ \ \ \  - 
 \biggl[{m^2\over H^2} +12 \xi -
 6\xi{\partial^2_t a \over a} - 
 12\xi \biggl({\partial_t a\over a}\biggr)^2\biggr]
 \phi^2 \biggr). 
 \label{B6}
\end{eqnarray}
Then using $\displaystyle \pi(t,\theta) =
{\partial {\cal L} \over \partial (\partial_t\phi)}$  one finds
\begin{equation}
\pi(t,\theta) =
Ha^{3-6\xi}\partial_t( a^{6\xi}\phi).
\end{equation}
Therefore, in terms of the creation and annihilation operators,
\begin{equation}
\pi(t,\theta) = Ha^{3-6\xi}\sum_{\{odd \ n\}}b_{(n)}
\partial_t( a^{6\xi}u_{(n)}(t,\theta)) +
b^\dagger_{(n)}\partial_t( a^{6\xi}u^*_{(n)}(t,\theta))\label{B7}
\end{equation}

The expressions \ref{B1} and \ref{B7} can be inverted to determine
$b_{(n)}$ in terms of $\pi(t,\theta)$ and $\phi(t,\theta)$. First observe that 
the linear combination of these fields satisfies the following relation:
\begin{eqnarray}
\int \biggl(
Ha^{3-6\xi}\partial_t(a^{6\xi}&&u^*_{(n)}(t,\theta))\phi(t,\theta)
-  u^*_{(n)}(t,\theta)\pi(t,\theta)\biggr)d\mu(c)
   = \sum_{\{odd \ n'\}}b_{(n')} \int 
F_{(n)(n')}(t,\theta)d\mu(c)\nonumber\\
F_{(n)(n')}(t,\theta)=&& Ha^{3-6\xi}\biggl(
\partial_t( a^{6\xi}u^*_{(n)}(t,\theta))u_{n'l'm'}(t,\theta)- 
u^*_{(n)}(t,\theta)\partial_t( a^{6\xi}u_{n'l'm'}(t,\theta))\biggr).
\nonumber
\end{eqnarray}
Next the inner product \ref{B3} can be used to reduce the right hand side 
of this equation to a delta function; thus
\begin{equation}
\int \biggl(
Ha^{3-6\xi}\partial_t(a^{6\xi}u^*_{(n)}(t,\theta))\phi(t,\theta)
-  u^*_{(n)}(t,\theta)\pi(t,\theta)\biggr)d\mu(c)
    = ib_{(n)}.\label{B8}
\end{equation}

Now the equal time commutation relations for the scalar field are
\[
\bigl[\phi(t,\theta),\pi(t,\theta')\bigr]=i\delta^3(\theta-\theta')
\ \ \ \ \ \ \ 
\bigl[\phi(t,\theta),\phi(t,\theta')\bigr]= 
\bigl[\pi(t,\theta),\pi(t,\theta')\bigr]=0
\]
where the delta function is a density of weight $-1/2$,
$\int  \delta^3(\theta-\theta') d\mu (c)= 1$.
The field representation of these operators can be found from these relations;
\begin{eqnarray}
\phi(t,\theta) &=& \sum_{\{odd \ n\}}\varphi_{(n)}Q_{(n)}(\theta)\nonumber\\
\pi(t,\theta) &=& \sum_{\{odd \ n\}}-iQ_{(n)}(\theta){d\ \over d\varphi_{(n)}}
\nonumber
\end{eqnarray}
where $\varphi_{(n)}$ are the  boundary data coefficients \ref{3.21} on the 
constant $t$ hypersurface. Using these expressions on the left hand side of 
\ref{B8} and observing that the hyperspherical harmonics  are orthogonal 
results in the desired field representation of $b_{(n)}$
\begin{equation}
b_{(n)}= -iHa^{3-6\xi}\partial_t\biggl(a^{6\xi}U^*_n(t)\biggr)\varphi_{(n)}
 +U^*_n(t){d\ \over d\varphi_{(n)}}.\label{B9}
\end{equation}

One can now construct the field representation of the vacuum 
${\cal Q}[RP^3;t,\varphi]$ by solving \ref{B4} in field representation,
\begin{equation}
b_{(n)}{\cal Q}[RP^3;t,\varphi] = 0
\end{equation}
where $b_{(n)}$ is given by \ref{B9}. This equation is first order for 
each value $\varphi_{(n)}$ and can be explicitly integrated. It follows that
\begin{eqnarray}
{\cal Q}[RP^3;t,\varphi]&\sim&\exp(iA(a,\varphi))\nonumber\\
A&=&{iH \over 2}a^3\sum_{\{ odd \ n\} }
{\partial_t(a^{6\xi} U^*_n(t))\over a^{6\xi} U^*_n(t)} \varphi^2_{(n)}.
\label{B10}
\end{eqnarray}
Changing variables to $\eta$, using 
$\displaystyle a(\eta) ={1 \over H\sin\eta}$
and the expression for $U^*_n(\eta)$ in terms of the modes \ref{3.20}, one 
finds
\begin{equation}
A={ia^2 \over 2}\sum_{\{ odd \ n\} } 
\biggr[{\partial_\eta(\sin^{\frac 12}\eta\rho_n(\eta)
 )\over 
 \sin^{\frac 12}\eta\rho_n(\eta)} -(6\xi-1)
 {\cos\eta}\biggr]
 \varphi^2_{(n)}.\label{B11}
\end{equation}
Evaluating $A$ at ${\eta=\eta_0}$ where $\displaystyle 
\eta_0=\sin^{-1} \frac 1{\cosh t_0} + \pi/2$ results in the classical action 
\ref{3.27}. Therefore \ref{B10} is exactly the same wavefunction as 
\ref{3.31} with the time coordinate $t_0$ related to the radius $a_0$ by the 
Lorentzian de Sitter solution. Thus the generalized Hartle-Hawking boundary 
condition produces the Bunch-Davies vacuum for the expanding $RP^3$ de Sitter 
extrema.

\begin{figure}
\begin{center}
\vglue 1 cm
\leavevmode
\epsfysize=14cm
\epsfbox{cone.epsf}
\end{center}
\vskip 1 cm
\caption{%
The construction of a cone of a topological space $V$. In this example, 
$V$ is a one dimensional space consisting of a figure eight as seen at upper 
left. The product space $V\times I$ is illustrated in the middle; it is a two
dimensional space consisting of two cylinders with points identified on the line 
where they touch. The cone of $V$ constructed from this product space is 
illustrated at  lower right; all points at $t=1$ are now identified to a single 
point.} 
\label{cone}
\end{figure}

\newpage

\begin{figure}
\begin{center}
\vglue 1 cm
\leavevmode
\epsfysize=14cm
\epsfbox{crp2.epsf}
\end{center}
\vskip 1 cm
\caption{%
The construction of $C(RP^2)$. In this construction, the space $RP^2$ 
is represented by a disk with antipodal points identified as seen at upper left. 
This representation and its connection with other representations of $RP^2$ is 
illustrated in Figure 3. The product space $RP^2\times I$ is illustrated 
in the middle. Note that  the front surface of this three dimensional  cylinder 
is identified with the back surface in accordance with the identification of 
the edges of  $RP^2$. Thus this  space is compact with two $RP^2$ boundaries. 
The cone $C(RP^2)$ is again formed by identifying all points at $t=1$ to one 
point as illustrated at lower right. Observe that $C(RP^2)$ has only 
one  boundary, that of the base $RP^2$.} 
\label{crp2}
\end{figure}

\newpage

\begin{figure}
\begin{center}
\vglue 1 cm
\leavevmode
\epsfysize=14cm
\epsfbox{rpn.epsf}
\end{center}
\vskip 1 cm
\caption{%
Constructions of $RP^n$. These constructions are explicitly illustrated 
for the case of $RP^2$. A familiar realization of $RP^n$  is given by  the 
identification of antipodal points on $S^n$; explicitly embedding $S^n$ in 
${\bf R}^{n+1}$, this identification is $x^a \to -x^a$ where $x^a$ is the 
cartesian coordinate of the point. The fundamental region of $RP^n$, that is 
the set of distinct points, is the upper hemisphere of $S^n$. The boundary 
of this region, the equator of the n-sphere, still has antipodal points 
identified; specifically, points on the front arc $a$ are identified 
with those on the back arc $a$ as indicated by the arrows. This 
representation  of $RP^n$ is  illustrated at lower left. As the hemisphere is 
homeomorphic to a n-ball, a topologically equivalent representation of $RP^n$ 
is given by a n-ball with antipodal points on its boundary identified  as
illustrated at lower right.} 
\label{rpn}
\end{figure}

\newpage

\begin{figure}
\begin{center}
\vglue 1 cm
\leavevmode
\epsfysize=14cm
\epsfbox{crp3.epsf}
\end{center}
\vskip 1 cm
\caption{%
A representation of C($RP^3$). 
This four dimensional conifold cannot 
be explicitly drawn, but can be represented as a sequence of three dimensional 
slices. Each slice is $RP^3$ represented as a solid 3-ball with antipodal 
points on the surface identified as illustrated at lower right in 
Figure 3.  The slice at $\tau=0$ consists of a single point.} 
\label{rp3}
\end{figure}

\end{document}